\documentclass[11pt,a4paper]{article}

\usepackage{amsmath}
\usepackage{amsthm}
\usepackage{amsfonts}
\usepackage{amssymb}
\usepackage{stmaryrd}
\usepackage{latexsym}
\usepackage{longtable}

\theoremstyle{definition}
\newtheorem{definition}{Definition}[section]
\newtheorem{remark}{Remark}[section]
\newtheorem{example}{Example}[section] 
\newtheorem{theorem}{Theorem}[section]

\newtheorem{corollary}{Corollary}[section]
\newtheorem{algorithm}{Algorithm}[section]

\begin{document}

{\title{An encryption algorithm using a generalization of the Markovski algorithm and a system of orthogonal
operations based on $T$-quasigroups}
	\author{{Nadezhda Malyutina, Alexander Popov, Victor Shcherbacov}
	}

		\maketitle

\begin{abstract}
Here is a more detailed description of the algorithm proposed in \cite{K1}. This algorithm simultaneously uses
two cryptographic procedures: encryption using a generalization of the Markovski algorithm \cite{K2} and
encryption using a system of orthogonal operations.
In this paper, we present an implementation of this algorithm based on $T$-quasigroups, more precisely, based on
medial quasigroups.
  			
		\medskip
		
		\noindent \textbf{2000 Mathematics Subject Classification:} 20N15, 20N05, 05B15, 94A60
		
		\medskip
		
		\noindent \textbf{Key words and phrases:} cipher, ciphertext, plaintext, Markovski algorithm,
quasigroup, $n$-ary groupoid, $T$-quasigroup.
	\end{abstract}
	
	\tolerance2500

\section{Basic concepts and definitions}

 This paper generalizes a cryptographic algorithm based on the use of quasigroups of a special form and provides
 an example illustrating the operation of this algorithm.

 Additional information is required for the modifications proposed in this work \cite{K3, K4}.

\begin{definition}
A binary groupoid $(Q, A)$ is a non-empty set $Q$ with a binary operation $A$ defined on it. An $n$-ary groupoid
$(Q, A)$ is a non-empty set $Q$ with an $Q$-ary operation $f$ defined on it.
\end{definition}

\begin{definition} $n$-ary groupoids $(Q,A_1),(Q,A_2),\dots,(Q,A_n)$ are called orthogonal if for any fixed
$n$-tuple $a_1,a_2,\dots,a_n$ the system of the following equations
		\[
	\left\{
	\begin{aligned}
	A_1(x_1,x_2,\dots,x_n)=a_1,\\
	A_2(x_1,x_2,\dots,x_n)=a_2,\\
		\ldots\\
			A_n(x_1,x_2,\dots,x_n)=a_n
	\end{aligned}
	\right.
	\]
	has a unique solution.
\end{definition}
If the set $Q$ is finite, then any system of $n$ orthogonal $n$-ary groupoids $(Q,A_i)$, $i\in \overline{1,n}$,
defines a permutation of the set $Q^n$ and vice versa \cite{K5, K6, K7}. Therefore, if $|Q| = q$, then there are
$(q^n)!$ systems of $n$-ary orthogonal groupoids defined on the set $Q$.

There are various generalizations of the definition of orthogonality of $n$-ary operations \cite{K8, K9}.
\begin{definition} A binary groupoid $(Q, A)$ is called a quasigroup if $\forall a, b \in Q$ two equations are
uniquely solvable:  $A(x,a)=b$  and $A(a,x)=b$.
\end{definition}
This definition of a quasigroup is called equational.
\begin{definition} Garret Birkhoff \cite{K10, K11}  defined an equational quasigroup as an algebra with three
binary operations $(Q,\cdot,\backslash,/)$ that satisfy the following six identities:
	\begin{center}
		$x\cdot (x\backslash y)=y$,\\
		$ (y/x)\cdot x=y$,\\
		$ x\backslash (x\cdot y)=y,$\\
		$ (y\cdot x)/ x=y$,\\
		$ x/(y\backslash x)=y,$\\
		$ (x/y)\backslash x=y.$
	\end{center}
	\end{definition}

\begin{definition} With a given binary quasigroup $(Q,A)$ there are $(3!-1)$ other, so-called parastrophes of the
quasigroup $(Q,A)$:
	\begin{center}
		$ A(x_1,x_2)=x_3$,\\
		$ A^{(12)}(x_2,x_1)=x_3$,\\
		$ A^{(13)}(x_3,x_2)=x_1$,\\
		$ A^{(23)}(x_1,x_3)=x_2$,\\
		$ A^{(123)}(x_2,x_3)=x_1$\\
		$ A^{(132)}(x_3,x_1)=x_2.$
	\end{center}
	\end{definition}
\begin{definition} Let $(Q,\cdot)$ be a groupoid and $a$ be a fixed element of $Q$, then we denote the left,
right, and middle translations as $L_a$, $R_a$, and $P_a$, respectively, and define them as follows:
	 $L_a x=a\cdot x$,   $R_a x=x\cdot a$,   	$x\cdot P_a x=a$, $\forall x \in Q$.
		\end{definition}
Translations play an important role in the theory of quasigroups.

In the following Table 1, for each type of translation, the equivalent parastrophe of the quasigroup $(Q,\cdot)$
is shown in each of the six cases \cite{K12}.
	\begin{longtable}{|c|c|c|c|c|c|c|}
	\caption{Translations of quasigroup parastrophes}
	\label{tab:table1}\\
	\hline
	 & $\epsilon$ & $(12)$ & $(13)$ & $(23)$ & $(123)$ & $(132)$ \\
	\hline
	$R$ & $R$ & $L$  & $R^{-1}$ & $P$ & $P^{-1}$ & $L^{-1}$ \\
	\hline
	$L$ & $L$ & $R$  & $P^{-1}$ & $L^{-1}$ & $R^{-1}$ & $P$  \\
	\hline
	$P$ & $P$ & $P^{-1}$  & $L^{-1}$ & $R$ & $L$ & $R^{-1}$ \\
	\hline
	$R^{-1}$ & $R^{-1}$ &$L^{-1}$  & $R$ & $P^{-1}$ & $P$ & $L$ \\
	\hline
	$L^{-1}$ & $L^{-1}$ &$R^{-1}$  & $P$ & $L$ & $R$ &$P^{-1}$ \\
	\hline
	$P^{-1}$ & $P^{-1}$ & $P$  & $L$ & $R^{-1}$ & $L^{-1}$ & $R$ \\
	\hline
	\end{longtable}
\begin{definition} A non-empty set $Q$  with an $n$-ary operation $f$ defined on it, such that in the equation
$f(x_1,x_2,\dots,x_n)=x_{n+1}$ the knowledge of any $n$ elements from $x_1,x_2,\dots, x_n,x_{n+1}$, makes it
possible to uniquely determine the remaining $(n+1)$\textit{th} element, is called an $n$-ary quasigroup.
\end{definition}

\begin{definition} An $n$-ary groupoid $(Q,f)$ is said to be isotopic to an $n$-ary groupoid $(Q,g)$ if there are
permutations $\mu_1,\mu_2,\dots,\mu_n$ of the set $Q$ such that:$f(x_1,x_2,\dots, x_n)=\mu^{-1} g(\mu_1 x_1,\mu_2
x_2,\dots,\mu_n x_n)$,$\forall x_1,x_2,\dots,x_n \in Q.$
	\end{definition}
It is written: $(Q,f)=(Q,g)T$, where $T=(\mu_1,\mu_2,\dots,\mu_n) - $ is an isotopy of an $n$- ary groupoid.

If $f=g$, then we get an autotopy of the $n$-ary groupoid $(Q,f)$. The last component of an autotopy is called a
quasiautomorphism.

If $\mu_1=\mu_2=\dots=\mu_n,$ then the groupoids $(Q,f)$ and $(Q,g)$ are said to be isomorphic.

And finally, if $\mu_1=\mu_2=\dots=\mu_n=\mu,$ and $f=g,$  then we obtain an automorphism of the groupoid
$(Q,f)$.

\begin{definition} $n$-ary quasigroup $(Q,f)$ of the form:
	  \begin{center}
	$f(x_{1}^{n})=\alpha_1 x_1+\alpha_2 x_2+\cdots +\alpha_n x_n+a=\sum_{i=1}^{n}\alpha_i x_i +a,$
	\end{center}
where $(Q,+)$ is a group, $\alpha_1,\alpha_2,\dots,\alpha_n$ are some automorphisms of the group $(Q,+),$ and an
element $a$ is some fixed element from the set $Q$, we will call a linear $n$-ary quasigroup (over the group
$(Q,+)$) \cite{K12}.
\end{definition}

\begin{definition} An $n$-ary linear quasigroup $(Q,f)$  over an Abelian group $(Q,+)$  is called an $n$-ary
$T$-quasigroup \cite{K13}. If $n=2$, then a quasigroup from this class of quasigroups is called a $T$- quasigroup
\cite{K14, K15}.
	
	The following identity for an $n$-ary quasigroup $(Q,f)$
	\begin{center}
		
$f(f(x_{11},x_{12},\dots,x_{1n}),f(x_{21},x_{22},\dots,x_{2n}),\dots,f(x_{n1},x_{n2},\dots,x_{nn}))=f(f(x_{11},x_{21},\dots,x_{n1}),f(x_{12},x_{22},\dots,x_{n2}),\dots,f(x_{1n},x_{2n},\dots,x_{nn}))$
	\end{center}
is called the medial identity \cite{K16}. An $n$-ary quasigroup with a medial identity is called an  $n$-ary
medial quasigroup.

In the binary case, we get the usual medial identity:
\begin{center}
	$xy\cdot uv=xu \cdot yv.$
\end{center}
\end{definition}

\begin{definition} A quasigroup $(Q,\cdot)$  is a $T$-quasigroup if and only if there exists an Abelian group
$(Q,+)$, its automorphisms $\phi$ and $\psi$, and a fixed element $c\in Q$ such that $x\cdot y = \phi x+\psi y
+c$, for all $x,y \in Q$ \cite{K14, K15}. A $T$-quasigroup with the additional condition  $\phi \psi=\psi \phi$ is
called medial.
\end{definition}

Medial quasigroups, like other classes of quasigroups isotopic to groups, allow one to construct quasigroups with
predetermined properties. Often these properties can be expressed in terms of the properties of groups and isotopy
components.

In the following theorem, the expression $A \perp {}^{(23)}A$ means that the quasigroups $(Q,A)$ and
$(Q,{}^{(23)}A)$ are orthogonal \cite{K12}.
\begin{theorem} For a finite quasigroup $(Q,A)$, the following equivalences hold:\\
	$(i)  A \perp {}^{(12)}A \Leftrightarrow ((x\backslash z) \cdot x=(y \backslash z) \cdot y \Rightarrow
x=y)$;\\
	$(ii)  A \perp {}^{(13)}A \Leftrightarrow (zx\cdot x=zy \cdot y \Rightarrow x=y)$;\\
	$(iii)  A \perp {}^{(23)}A \Leftrightarrow (x \cdot xz=y \cdot yz \Rightarrow x=y)$;\\
	$(iv) A \perp {}^{(123)}A \Leftrightarrow (x \cdot zx=y \cdot zy\Rightarrow x=y)$;\\
	$(v)  A \perp {}^{(132)}A \Leftrightarrow (xz \cdot x=yz \cdot y \Rightarrow x=y)$, for all $x,y,z \in Q$.
\end{theorem}
To construct the quasigroups mentioned in Theorem 2.1, one can use GAP and Prover \cite{K17}.

If $(Q,\cdot)$ is a $T$-quasigroup of the form $x \cdot y = \phi x+\psi y+c$, then its parastrophes have the
following forms, respectively:
\begin{center}
$x\overset{(12)}{\cdot}y=\psi x+\phi y+c,$\\
$x\overset{(13)}{\cdot}y=\phi^{-1} x-\phi^{-1} \psi y-\phi^{-1} c,$\\
$x\overset{(23)}{\cdot}y=-\psi^{-1} \phi x+\psi^{-1} y-\psi^{-1} c,$\\
$x\overset{(123)}{\cdot}y=-\phi^{-1} \psi x+\phi^{-1} y-\phi^{-1} c,$\\
$x\overset{(132)}{\cdot}y=\psi^{-1} x-\psi^{-1} \phi y-\psi^{-1} c.$\\
\end{center}

To construct a quasigroup $(Q,A)$ orthogonal to its parastrophe in a more theoretical way, one can use the
following theorem \cite{K4,K18}.
\begin{theorem} For a $T$-quasigroup $(Q,A)$ of the form: $A(x,y) = \phi x + \psi y + c,$  the following
equivalences hold over an abelian group $(Q,+)$:\\
	$(i)  A \perp {}^{(12)}A \Leftrightarrow (\phi -\psi),(\phi+\psi)$) are permutations of the set $Q;$\\
	$(ii)  A \perp {}^{(13)}A \Leftrightarrow (\epsilon+\phi)$is a permutation of the set $Q$;\\
	$(iii)  A \perp {}^{(23)}A \Leftrightarrow (\epsilon+\psi)$is a permutation of the set $Q$;\\
	$(iv) A \perp {}^{(123)}A \Leftrightarrow (\phi+\psi^{2})$is a permutation of the set $Q$;\\
	$(v)  A \perp {}^{(132)}A \Leftrightarrow (\phi^{2}+\psi)$is a permutation of the set $Q$.
\end{theorem}
\begin{corollary} $T$-quasigroup $(Z_p,\circ)$ of the form $x\circ y=k\cdot x+m\cdot y+c$, where $(Z_p,+)$  is a
cyclic group of prime order $p,k,m,c \in Z_p;$ $k,m,k + m$,$k - m,k + 1,m + 1, k^2+ m,k + m^2\not \equiv 0 \pmod
{p}$, where the operation $"\cdot$" is multiplication modulo $p$, is orthogonal to any of its parastrophe.
\end{corollary}
	Quasigroups from Corollary 1 are suitable objects for constructing the above algorithms \cite{K4}.
	
Belousov \cite{K19} has a proof of Toyoda's theorem, namely: if $(Q,A)$ is a medial quasigroup, then there exists
an Abelian group $(Q,+)$ such that the operation A has the form: $A(x,y) = \phi x + \psi y + c$, where $\phi,
\psi$ are automorphisms of the group $(Q,+)$, moreover, $\phi\psi= \psi\phi$, and $c$ is some fixed element of
$Q$.

The converse Toyoda theorem also holds.
\begin{theorem} (Toyoda's inverse theorem) Let $(Q,+)$ be an arbitrary abelian group, $\phi, \psi$ be
automorphisms of the group $(Q,+)$, where $\phi\psi= \psi\phi$, $c \in Q$ is a fixed arbitrary element. Then
$(Q,\cdot)$ is a medial quasigroup, where $x \cdot y = \phi x + \psi y + c$,$\forall x,y \in Q$.
\end{theorem}
\section{Results}
Below we denote the action of the left (right, middle) translation to the power of $a$ of a binary quasigroup
$(Q,g_1)$ on an element $u_1$ by the symbol  ${}_{g_1} T{}_{l_1}^a (u_1)$. And so on. Here  $l_1$ means the leader
element.

A description of Algorithm 1 and an example that implements the operation of this algorithm can be found in
\cite{K3}. Now we give a generalization of Algorithm 1. Note that the algorithm works for texts with even length.
\begin{algorithm}
	\textbf {(Algorithm 1*).}\\
	\upshape \textbf {Encryption.} Initially, we have the plaintext $u_1,u_2,\dots,u_{2n}$.
	\begin {center}
	{\bfseries Step 1.}\\
	${}_{g_1} T{}_{l_1}^{a_{11}}(u_1)=v_1,$ ${}_{g_2} T{}_{l_2}^{a_{12}}(u_2)=v_2,$			
$F_1^{a_{13}}(v_1,v_2)=(v_1^{'},v_2^{'});$\\
			
		{\bfseries Step 2.}\\
			${}_{g_3} T{}_{v_1^{'}}^{a_{21}}(u_3)=v_3,$	${}_{g_4} T{}_{v_2^{'}}^{a_{22}}(u_4)=v_4,$
				$F_2^{a_{23}}(v_3,v_4)=(v_3^{'},v_4^{'});$\\
				
			{\bfseries Step 3.}\\
				${}_{g_5} T{}_{v_3^{'}}^{a_{31}}(u_5)=v_5,$	${}_{g_6} T{}_{v_4^{'}}^{a_{32}}(u_6)=v_6,$		
$F_3^{a_{33}}(v_5,v_6)=(v_5^{'},v_6^{'});$\\
			$	\cdots $\\
			{\bfseries Step N.}\\
						${}_{g_{2n-1}}T{}_{v_{2n-1}^{'}}^{a_{n_1}}(u_{2n-1})=v_{2n-1},$						
${}_{g_{2n}} T{}_{v_{2n}^{'}}^{a_{n_2}}(u_{2n})=v_{2n},$							
$F_n^{a_{n_3}}(v_{2n-1},v_{2n})=(v_{2n-1}^{'},v_{2n}^{'}).$\\
	\end {center}

We get the ciphertext $v_1^{'},v_2^{'},\dots,v_{2n}^{'}$.

\normalfont {Thus, to encrypt a plaintext of length $2n$, we need $2n$ different operations on the quasigroups
used in Algorithm 1*, $n$ different values of the function $F$, $2$ leader elements, and $3n$ different powers
used in the translation algorithm.}\\
 {\textbf {Decryption.}}  {Let us have a ciphertext of the form  $v_1^{'},v_2^{'},\dots,v_{2n}^{'}$.}

  	\begin {center}
  	
	{\bfseries Step 1.}\\
	$F_1^{-a_{13}}(v_1^{'},v_2^{'})=(v_1,v_2),$	${}_{g_1} T{}_{l_1}^{-a_{11}}(v_1)=u_1,$
	${}_{g_2} T{}_{l_2}^{-a_{12}}(v_2)=u_2;$\\
	{\bfseries Step 2.}\\
	$F_2^{-a_{23}}(v_3{'},v_4{'})=(v_3,v_4);$	${}_{g_3} T{}_{v_1^{'}}^{-a_{21}}(v_3)=u_3,$
	${}_{g_4} T{}_{v_2^{'}}^{-a_{22}}(v_4)=u_4,$\\
		
   	{\bfseries Step 3.}\\
   	$F_3^{-a_{33}}(v_5^{'},v_6^{'})=(v_5,v_6);$
	${}_{g_5} T{}_{v_3^{'}}^{-a_{31}}(v_5)=u_5,$
	${}_{g_6} T{}_{v_4^{'}}^{-a_{32}}(v_6)=u_6,$\\
	$	\cdots $\\
	{\bfseries Step N.}\\
		$F_n^{-a_{n_3}}(v_{2n-1}^{'},v_{2n}^{'})=(v_{2n-1},v_{2n}).$\\
	${}_{g_{2n-1}} T{}_{v_{2n-1}^{'}}^{-a_{n_1}}(v_{2n-1})=u_{2n-1},$
	${}_{g_{2n}} T{}_{v_{2n}^{'}}^{-a_{n_2}}(v_{2n})=u_{2n},$\\
	\end {center}
	We received a plaintext of the form $u_1,u_2,\dots,u_{2n}$.
	\end{algorithm}
From Algorithm 1* we obtain the classical Markovski algorithm if we take only one quasigroup, one kind of
quasigroup translation (left translations), each of which is taken to the first degree, and if the system of
orthogonal operations (crypto procedures $F$) is not used. Some generalizations of Algorithm 1 are given in
\cite{K4}.

The Markovski algorithm is a special case of Algorithm 1*. As in the Markovski algorithm, in Algorithm 1*, the
powers $a_{11},a_{12}, \dots,a_{n3}$ must be different to protect this algorithm from chosen plaintext and
ciphertext attacks \cite{K3}.

Consider the operation of Algorithm 1* based on $T$-quasigroups using a specific example.
\begin{example}
Take a cyclic group $(Z_{313},+) = (A,+)$.

Initially, we have a plaintext $u_1,u_2,u_3,u_4,u_5,u_6$ of length $k=6$.

$1)$ First, we define all the $T$-quasigroups we need.\\
$1.$ Define a $T$-quasigroup $(A,g_1)=(A,*)$  of the form:
\begin{center}
 $x*y=25 \cdot x+37 \cdot y+11$\\
 \end{center}
  with leader $l_1$. The mapping $x \to x *l_1$ will be denoted by $R_{l_1}$,i.e., $R_{l_1} (x)=x*l_1$ for all $x
  \in A$.

To find the mapping $R_{l_1}^{-1}$ taking into account Table 1, we find the type of operation $\overset{(13)}{*}$
using the formula:
$x\overset{(13)}{*}y=\phi^{-1} x-\phi^{-1} \psi y-\phi^{-1} c.$
 \begin{center}
$\phi=25, \psi=37, c=11$,\\
$ \phi^{-1}\equiv 25^{311}(\mod 313) \equiv 288, -\phi^{-1}\equiv 25$,\\
$ -\phi^{-1} \psi \equiv 25 \cdot 37(\mod313) \equiv 299$,\\
$ -\phi^{-1} c \equiv 25 \cdot 11(\mod313) \equiv 275$.\\
\end{center}
We have:
\begin{center}
	 $x\overset{(13)}{*}y=288 \cdot x+299 \cdot y+275,$
\end{center}
$ R_{l_1}^{-1}=x\overset{(13)}{*}l_1=R_{l_1}^{(13)}x.$
In a sense, the quasigroup $(A,\overset{(13)}{*})$ is a “right inverse quasigroup” to the quasigroup $(A,*)$. Note
that Corollary 2.1 implies that $(A,*) \perp (A,\overset{(13)}{*})$.\\
$2.$ Define a $T$-quasigroup $(A,g_2)=(A,\circ)$  of the form:
\begin{center}
	$x\circ y=75 \cdot x+39 \cdot y+100$
\end{center}
with leader $l_2$. Mapping $x \to l_2\circ x$ is denoted by $L_{l_2}$,i.e., $L_{l_2} (x)=l_2 \circ x$ for all $x
\in A$.

To find the mapping $L_{l_2}^{-1}$, we use Table 1 and find the type of operation $\overset{(23)}{\circ}$ by the
formula:
$x \overset{(23)}{\circ}y=-\psi^{-1} \phi x+\psi^{-1} y-\psi^{-1} c$.
\begin{center}
	$\phi=75, \psi=39, c=100$,\\
	$ \psi^{-1}\equiv 39^{311}(\mod 313) \equiv 305, -\psi^{-1}\equiv 8$,\\
	$ -\psi^{-1} \phi \equiv 8 \cdot 75(\mod313) \equiv 287$,\\
	$ -\psi^{-1} c \equiv 8 \cdot 100(\mod313) \equiv 174$.
\end{center}

We have:
\begin{center} $x\overset{(23)}{\circ}y=287 \cdot x+305 \cdot y+174.$\\
	\end{center}
$3.$ Let us define a $T$-quasigroup $(A,g_3)=(A,\star)$  of the form:
\begin{center}
	$x\star y=127 \cdot x+213 \cdot y+19$.
\end{center}
Mapping $x \to y\star x$ is denoted by $L_y$,i.e., $L_y(x)=y\star x$ for all $x \in A$.

To find the mapping $L_y^{-1}$, use Table 1 and find the type of operation $\overset{(23)}{\star}$ by the
formula:
$x \overset{(23)}{\star}y=-\psi^{-1} \phi x+\psi^{-1} y-\psi^{-1} c$.
\begin{center}
	$\phi=127, \psi=213, c=19$,\\
	$ \psi^{-1}\equiv 213^{311}(\mod 313) \equiv 241, -\psi^{-1}\equiv 72$,\\
	$ -\psi^{-1} \phi \equiv 72 \cdot 127(\mod313) \equiv 67$,\\
	$ -\psi^{-1} c \equiv 72 \cdot 19(\mod313) \equiv 116$.
\end{center}

We have:
\begin{center}
	 $x\overset{(23)}{\star}y=67 \cdot x+241 \cdot y+116.$ \\
	 \end{center}
$4.$ Let us define a $T$-quasigroup $(A,g_4)=(A, \diamond)$  of the form:
\begin{center}
	$x\diamond y=151 \cdot x+301 \cdot y+199$.
\end{center}
Denote the mapping  $x \to x\diamond y$ by $R_y$,i.e.,  $R_y(x)=x \diamond y$ for all $x \in A$.

To find the mapping $R_y^{-1}$, taking into account Table 1, we find the type of operation
$\overset{(13)}{\diamond}$ using the formula:
$x \overset{(13)}{\diamond}y=\phi^{-1}x-\phi^{-1}\psi y-\phi^{-1} c$.
\begin{center}
	$\phi=151, \psi=301, c=199$,\\
	$ \phi^{-1}\equiv 151^{311}(\mod 313) \equiv 199, -\phi^{-1}\equiv 114$,\\
	$ -\phi^{-1} \psi \equiv 114 \cdot 301(\mod313) \equiv 197$,\\
	$ -\phi^{-1} c \equiv 114 \cdot 199(\mod313) \equiv 150$.
\end{center}

We have:
\begin{center} $x\overset{(23)}{\diamond}y=199 \cdot x+197 \cdot y+150.$\\
\end{center}
$5.$ Let us define a $T$-quasigroup $(A,g_5)=(A,\odot)$  of the form:
\begin{center}
	$x\odot y=213 \cdot x+3 \cdot y+9$.
\end{center}
The mapping $x \to x\odot y$ will be denoted by $R_y$,i.e., $R_y(x)=x \odot y$ for all $x \in A$.

To find the mapping $R_y^{-1}$, taking into account Table 1, we find the type of operation $\overset{(13)}{\odot}$
using the formula:
$x \overset{(13)}{\odot}y=\phi^{-1}x-\phi^{-1}\psi y-\phi^{-1} c$.
\begin{center}
	$\phi=213, \psi=3, c=9$,\\
	$ \phi^{-1}\equiv 213^{311}(\mod 313) \equiv 241, -\phi^{-1}\equiv 72$,\\
	$ -\phi^{-1} \psi \equiv 72 \cdot 3(\mod313) \equiv 216$,\\
	$ -\phi^{-1} c \equiv 72 \cdot 9(\mod313) \equiv 22$.
\end{center}

We have:
\begin{center} $x\overset{(23)}{\odot}y=241 \cdot x+216 \cdot y+22.$\\
	\end{center}
$6.$ Let us define a $T$-quasigroup $(A,g_6)=(A, \oplus )$  of the form:
\begin{center}
	$x\oplus y=303 \cdot x+200 \cdot y+99$.
\end{center}
Mapping $x \to y\oplus x$ is denoted by $L_y$,i.e., $L_y(x)=y \oplus x$ for all $x \in A$.

To find the mapping $L_y^{-1}$, we use Table 1 and find the type of operation $\overset{(23)}{\oplus}$ by the
formula:
$x \overset{(23)}{\oplus}y=-\psi^{-1} \phi x+\psi^{-1} y-\psi^{-1} c$.
\begin{center}
	$\phi=303, \psi=200, c=99$,\\
	$ \psi^{-1}\equiv 200^{311}(\mod 313) \equiv 36, -\psi^{-1}\equiv 277$,\\
	$ -\psi^{-1} \phi \equiv 277 \cdot 303(\mod313) \equiv 47$,\\
	$ -\psi^{-1} c \equiv 277 \cdot 99(\mod313) \equiv 192$.
\end{center}
We have:
\begin{center} $x\overset{(23)}{\oplus}y=47 \cdot x+36 \cdot y+192.$
	\end{center}
  $ 2)$ We define systems of two parastrophic orthogonal $T$-quasigroups. We need $3$ systems of cryptofunctions
  $F$.\\
$1.$ We define a system of two parastrophic orthogonal $T$-quasigroups $(A,g_7 )= (A,\oslash)$ and
$(A,\overset{(23)}{\oslash})$  as follows:
\[
\left\{
\begin{aligned}
x\oslash y=7 \cdot x+12 \cdot y+13\\
x\overset{(23)}{\oslash} y=182 \cdot x+287 \cdot y+25.\\
\end{aligned}
\right.
\]

Denote the quasigroup system $(A,\oslash,\overset{(23)}{\oslash})$ by $F_1 (x,y)$, since this system is a function
of two variables. To find the mapping $F_1^{-1} (x,y)$, we solve the system of linear equations:
\begin{equation*}
\begin{cases}
7 \cdot x+12 \cdot y+13=a\\
182 \cdot x+287 \cdot y+25=b
\end{cases}
\Leftrightarrow
\begin{cases}
175y=182 \cdot a+306 \cdot b\\
138x=287 \cdot a+301 \cdot b+12
\end{cases}
\Rightarrow
\end{equation*}
\begin{equation*}
 F_1^{-1} (x,y):
\begin{cases}
x=86 \cdot a+136 \cdot b+177\\
y=289 \cdot a+25 \cdot b.
\end{cases}
\end{equation*}

Therefore, we have, if $F_1 (x,y)= (a,b)=(7 \cdot x+12 \cdot y+13 ,182 \cdot x+287 \cdot y+25)$, then  $F_1^{-1}
(a,b)= (86 \cdot a+136 \cdot b+177 ,289 \cdot a+25 \cdot b)$.\\
$2.$ We define a system of two parastrophic orthogonal $T$-quasigroups $(A,g_8 )= (A,\circledcirc)$ and
$(A,\overset{(23)}{\circledcirc})$  as follows:
\[
\left\{
\begin{aligned}
x\circledcirc y=79 \cdot x+113 \cdot y+23\\
x\overset{(23)}{\circledcirc} y=27 \cdot x+277 \cdot y+202.\\
\end{aligned}
\right.
\]

Denote the quasigroup system $(A,\circledcirc,\overset{(23)}{\circledcirc})$ by $F_2 (x,y)$, since this system is
a function of two variables. To find the mapping $F_2^{-1} (x,y)$, we solve the system of linear equations:
\begin{equation*}
\begin{cases}
79 \cdot x+113 \cdot y+23=a\\
27 \cdot x+277 \cdot y+202=b
\end{cases}
\Leftrightarrow
\begin{cases}
52y=-27 \cdot a+79 \cdot b\\
52x=277 \cdot a-113 \cdot b+179
\end{cases}
\Rightarrow
\end{equation*}
\begin{equation*}
F_2^{-1} (x,y):
\begin{cases}
x=216 \cdot a+52 \cdot b+178\\
y=162 \cdot a+152 \cdot b.
\end{cases}
\end{equation*}

Therefore, we have, if $F_2 (x,y)= (79 \cdot x+113 \cdot y+23 ,27 \cdot x+277 \cdot y+202)$, then  $F_2^{-1}
(a,b)= (216 \cdot a+52 \cdot b+178 ,162 \cdot a+152 \cdot b)$.\\
$3.$ We define a system of two parastrophic orthogonal $T$-quasigroups $(A,g_9 )= (A,\odot)$ and
$(A,\overset{(13)}{\odot})$  as follows:
\[
\left\{
\begin{aligned}
x\odot y=81 \cdot x+101 \cdot y+99\\
x\overset{(13)}{\odot} y=228 \cdot x+134 \cdot y+277.\\
\end{aligned}
\right.
\]

Denote the quasigroup system $(A,\odot,\overset{(13)}{\odot})$ by $F_3 (x,y)$, since this system is a function
of two variables. To find the mapping $F_3^{-1} (x,y)$, we solve the system of linear equations:
\begin{equation*}
\begin{cases}
81 \cdot x+101 \cdot y+99=a\\
228 \cdot x+134 \cdot y+277=b
\end{cases}
\Leftrightarrow
\begin{cases}
280y=228 \cdot a-81 \cdot b-135\\
280x=-134 \cdot a+101 \cdot b
\end{cases}
\Rightarrow
\end{equation*}
\begin{equation*}
F_3^{-1} (x,y):
\begin{cases}
x=42 \cdot a+272 \cdot b\\
y=50 \cdot a+287 \cdot b+61.
\end{cases}
\end{equation*}

Therefore, we have, if $F_3 (x,y)= (81 \cdot x+101 \cdot y+99 ,228 \cdot x+134 \cdot y+277)$, then  $F_3^{-1}
(a,b)= (42 \cdot a+272 \cdot b ,50 \cdot a+287 \cdot b+61)$.

Now we can use all the operations described above in Algorithm 1*.

Let the plaintext look like \textbf {56; 43; 105; 59; 61; 19}.

In the algorithm, we will use the following values of the degrees:\\
$a_{11}=3; a_{12}=1;a_{13}=2;a_{21}=1;a_{22}=3;a_{23}=2;a_{31}=3;a_{32}=2;a_{33}=1$.

Let us choose the elements as leaders: $l_1=110; l_2=210$.

\textbf {Encryption.}
\begin {center}
{\bfseries Step 1.}\\
${}_{g_1}T_{l_1}^{a_{11}}(u_1)= {}_{g_1} T_{110}^3 (56)=v_1,$\\
${}_{g_1} T_{110}^1(56)=u_1*l_1=56*110=25 \cdot 56+37 \cdot 110+11=160,$\\
${}_{g_1} T_{110}^2(56)={}_{g_1}T_{110}^1(160)=160*110=25 \cdot 160+37 \cdot 110+11=256,$\\
${}_{g_1} T_{110}^3(56)={}_{g_1} T_{110}^1(256)=256*110=25 \cdot 256+37 \cdot 110+11=152=v_1,$\\
${}_{g_2} T{}_{l_2}^{a_{12}}(u_2)={}_{g_2} T_{210}^1(43)=v_2,$\\
${}_{g_2} T{}_{210}^1(43)=l_2 \circ 43=75 \cdot 210+39 \cdot 43+100=312=v_2,$\\
$F_1^{a_{13}}(v_1,v_2)=F_1^2(152,312)=(v_1^{'},v_2^{'}),$\\
$F_1^1(152,312)=(7 \cdot 152+12 \cdot 312+13, 182 \cdot 152+287 \cdot 312+25)=(126,171),$
$F_1^2(152,312)=F_1^1(126,171)=(7 \cdot 126+12 \cdot 171+13, 182 \cdot 126+287 \cdot
171+25)=(130,44)=(v_1^{'},v_2^{'}),$

{\bfseries Step 2.}\\
${}_{g_3}T_{v_1^{'}}^{a_{21}}(u_3)= {}_{g_3} T_{130}^1 (105)=v_3,$\\
${}_{g_3} T_{130}^1(105)=130 \star 105=127 \cdot 130+213 \cdot 105+19=82=v_3,$\\
${}_{g_4}T_{v_2^{'}}^{a_{22}}(u_4)= {}_{g_4} T_{44}^3 (59)=v_4,$\\
${}_{g_4} T_{44}^1(59)=59 \lozenge 44=151 \cdot 59+301 \cdot 44+199=129,$\\
${}_{g_4} T_{44}^2(59)={}_{g_4} T_{44}^1(129)=129 \lozenge 44=151 \cdot 129+301 \cdot 44+199=57,$\\
${}_{g_4} T_{44}^3(59)={}_{g_4} T_{44}^1(57)=57 \lozenge 44=151 \cdot 57+301 \cdot 44+199=140=v_4,$\\
$F_2^{a_{23}}(v_3,v_4)=F_2^2(82,140)=(v_3^{'},v_4^{'}),$\\
$F_2^1(82,140)=(79 \cdot 82+113 \cdot 140+23, 27 \cdot 82+277 \cdot 140+202)=(98,193),$
$F_2^2(82,140)=F_2^1(98,193)=(79 \cdot 98+113 \cdot 193+23, 27 \cdot 98+277 \cdot
193+202)=(152,282)=(v_3^{'},v_4^{'}),$

{\bfseries Step 3.}\\
${}_{g_5}T_{v_3^{'}}^{a_{31}}(u_5)= {}_{g_5} T_{152}^3 (61)=v_5,$\\
${}_{g_5} T_{152}^1(61)=61 \odot 152=213 \cdot 61+3 \cdot 152+9=312,$\\
${}_{g_5} T_{152}^2(61)={}_{g_5}T_{152}^1(312)=312 \odot 152=213 \cdot 312+3 \cdot 152+9=252,$\\
${}_{g_5} T_{152}^3(61)={}_{g_5} T_{152}^1(252)=252 \odot 152=213 \cdot 252+3 \cdot 152+9=305=v_5,$\\
${}_{g_6} T{}_{v_4^{'}}^{a_{32}}(u_6)={}_{g_6} T_{282}^2(19)=v_6,$\\
${}_{g_6} T{}_{282}^1(19)=282 \circledast 19=303 \cdot 282+200 \cdot 19+99=140,$\\
${}_{g_6} T{}_{282}^2(19)={}_{g_6} T{}_{282}^1(140)=282 \circledast 140=303 \cdot 282+200 \cdot
140+99=239=v_6,$\\
$F_3^{a_{33}}(v_5,v_6)=F_3^1(305,239)=(v_5^{'},v_6^{'}),$\\
$F_3^1(305,239)=(81 \cdot 305+101 \cdot 239+99, 228 \cdot 305+134 \cdot 239+277)=(115,118)=(v_5^{'},v_6^{'}).$\\
\end {center}

We get the following ciphertext: \textbf {130; 44; 152; 282; 115; 118}.

{\textbf {Decryption.}}
\begin {center}
	{\bfseries Step 1.}\\
	$F_1^{-a_{13}}(v_1^{'},v_2^{'})=F_1^{-2}(130,44)=(v_1,v_2),$\\
	$F_1^{-1}(130,144)=(86 \cdot 130+136 \cdot 44+177, 289 \cdot 130+25 \cdot 44)=(126,171),$\\
	$F_1^{-2}(130,144)=F_1^{-1}(126,171)=(86 \cdot 126+136 \cdot 171+177, 289 \cdot 126+25 \cdot
171)=(152,312)=(v_1,v_2),$\\
	${}_{g_1} T{}_{l_1}^{-a_{11}}(v_1)={}_{g_1} T_{110}^{-3}(152)=u_1,$\\
	${}_{g_1} T_{110}^{-1}(152)=152\overset{(13)}{*} 110= 288 \cdot 152+299 \cdot 110+275=256,$\\
	${}_{g_1} T_{110}^{-2}(152)={}_{g_1} T_{110}^{-1}(256)=256\overset{(13)}{*} 110= 288 \cdot 256+299 \cdot
110+275=160,$\\
	${}_{g_1} T_{110}^{-3}(152)={}_{g_1} T_{110}^{-1}(160)=160\overset{(13)}{*} 110= 288 \cdot 160+299 \cdot
110+275=56=u_1,$\\
	${}_{g_2} T{}_{l_2}^{-a_{12}}(v_2)={}_{g_2} T_{210}^{-1}(312)=u_2,$\\
	${}_{g_2} T_{210}^{-1}(312)=210\overset{(23)}{\circ} 312= 287 \cdot 210+305 \cdot 312+174=43=u_2,$\\
	
	{\bfseries Step 2.}\\
$F_2^{-a_{23}}(v_3^{'},v_4^{'})=F_2^{-2}(152,282)=(v_3,v_4),$\\
$F_2^{-1}(152,282)=(216 \cdot 152+52 \cdot 282+178, 162 \cdot 152+152 \cdot 282)=(98,193),$\\
$F_2^{-2}(152,282)=F_2^{-1}(98,193)=(216 \cdot 98+52 \cdot 193+178, 162 \cdot 98+152 \cdot
193)=(82,140)=(v_3,v_4),$\\
${}_{g_3} T{}_{v_1^{'}}^{-a_{21}}(v_3)={}_{g_3} T_{130}^{-1}(82)=u_3,$\\
${}_{g_3} T_{130}^{-1}(82)=130\overset{(23)}{\star} 82= 67 \cdot 130+241 \cdot 82+116=105=u_3,$\\
${}_{g_4} T{}_{v_2^{'}}^{-a_{22}}(v_4)={}_{g_4} T_{44}^{-3}(140)=u_4,$\\
${}_{g_4} T_{44}^{-1}(140)=140\overset{(13)}{\lozenge} 44= 199 \cdot 140+197 \cdot 44+150=57,$\\
${}_{g_4} T_{44}^{-2}(140)={}_{g_4} T_{44}^{-1}(57)=57\overset{(13)}{\lozenge} 44= 199 \cdot 57+197 \cdot
44+150=129,$\\
${}_{g_4} T_{44}^{-3}(140)={}_{g_4} T_{44}^{-1}(129)=129\overset{(13)}{\lozenge} 44= 199 \cdot 129+197 \cdot
44+150=59=u_4,$\\

	{\bfseries Step 3.}\\
	$F_3^{-a_{33}}(v_5^{'},v_6^{'})=F_3^{-2}(115,118)=(v_5,v_6),$\\
	$F_3^{-1}(115,118)=(42 \cdot 115+272 \cdot 118, 50 \cdot 115+287 \cdot 118+61)=(305,239)=(v_5,v_6),$\\
	${}_{g_5} T{}_{v_3^{'}}^{-a_{31}}(v_5)={}_{g_5} T_{152}^{-3}(305)=u_5,$\\
	${}_{g_5} T_{152}^{-1}(305)=305\overset{(13)}{\circledcirc} 152= 241 \cdot 305+216 \cdot 152+22=252,$\\
	${}_{g_5} T_{152}^{-2}(305)={}_{g_5} T_{152}^{-1}(252)=252\overset{(13)}{\circledcirc} 152= 241 \cdot 252+216
\cdot 152+22=312,$\\
	${}_{g_5} T_{152}^{-3}(305)={}_{g_5} T_{152}^{-1}(312)=312\overset{(13)}{\circledcirc} 152= 241 \cdot 312+216
\cdot 152+22=61=u_5,$\\
	${}_{g_6} T{}_{v_4^{'}}^{-a_{32}}(v_6)={}_{g_6} T_{282}^{-2}(239)=u_6,$\\
	${}_{g_6} T_{282}^{-1}(239)=282\overset{(23)}{\circledast} 239= 47 \cdot 282+36 \cdot 239+192=140,$\\
	${}_{g_6} T_{282}^{-2}(239)={}_{g_6} T_{282}^{-1}(140)=282\overset{(23)}{\circledast} 140= 47 \cdot 282+36
\cdot 140+192=19=u_6.$\\
	\end {center}
	
We get the following plaintext: \textbf {56; 43; 105; 59; 61; 19}.\\
\end{example}

\section{Conclusion}

A program has been developed that uses a free version of the Pascal ABC programming language. The conducted
experiments show that encoding/decoding is performed quite quickly. The program works for any values of leaders,
powers, and any plaintext of length $6$. It can be easily modified for text of any length and any used quasigroups
and functions.

Algorithm 1* makes it possible to obtain an almost "natural" stream cipher, i.e., a stream cipher that encodes a
pair of plaintext elements at any step. It is easy to see that Algorithm 1* can be generalized to the $n$-ary
case.\\

\begin{remark}
Proper binary groupoids are preferable to linear quasigroups in terms of the construction of the map $F(x,y)$ for
greater encryption security, but in this case, decryption may be slower than in the case of linear quasigroups,
and the definition of these groupoids requires more memory. The same remark is valid for the choice of the
function $g$. Perhaps the golden mean in this choice problem is the use of linear quasigroups over non-Abelian,
especially simple, groups.	
\end{remark}

\begin{remark}
	In this cipher, there is a possibility of protection against the standard statistical attack. For this area,
more commonly used letters or pairs of letters can be denoted by more than one integer or more than one pair of
integers \cite{K3}.
\end{remark}

\vspace{2mm}

\begin{center}
\begin{parbox}{120mm}{\footnotesize
Nadeghda Malyutina$^{1}$, Alexander Popov$^{2}$, Victor Shcherbacov$^{3}$

\vspace{3mm}

\noindent $^{1}$ {Docent}\\
Tiraspol State University

\noindent E--mail: \texttt{231003.bab.nadezhda@mail.ru}

\vspace{3mm}

\noindent $^{2}$ {Student}\\
Tiraspol State University

\noindent E--mail: \texttt{hhhhhggg5dhn@gmail.com}

\vspace{3mm}

\noindent $^{3}$ {Col. st. pr.}

Moldova State University, Vladimir Andrunachievici Institute of Mathematics and Computer Science

\noindent E--mail: \texttt{vscerb@gmail.com}
}
\end{parbox}
\end{center}

\end{document}